\documentclass{emulateapj}

\def\gsim{\;\rlap{\lower 2.5pt
 \hbox{$\sim$}}\raise 1.5pt\hbox{$>$}\;}
\def\lsim{\;\rlap{\lower 2.5pt
   \hbox{$\sim$}}\raise 1.5pt\hbox{$<$}\;}
\def\micron{~$\mu\textrm{m}$ }
\def\micronend{$\mu\textrm{m}$}
\def\microjy{~$\mu\textrm{Jy}$ }
\def\microjyend{~$\mu\textrm{Jy}$}
\def\lengthunits{$h^{-1}\, \textrm{kpc}$ }

\begin{document}

\title{Evidence for a Population of High--Redshift Submillimeter Galaxies from Interferometric Imaging}
\author{Joshua D. Younger \altaffilmark{1,2}, Giovanni G. Fazio\altaffilmark{1}, Jia--Sheng Huang\altaffilmark{1}, Min S. Yun\altaffilmark{3}, Grant W. Wilson\altaffilmark{3}, Matthew L. N. Ashby\altaffilmark{1}, Mark A. Gurwell\altaffilmark{1}, Kamson Lai\altaffilmark{1}, Alison B. Peck\altaffilmark{1,4}, Glen R. Petitpas\altaffilmark{1}, David J. Wilner\altaffilmark{1}, Daisuke Iono\altaffilmark{5}, Kotaro Kohno\altaffilmark{6}, Ryohei Kawabe\altaffilmark{7}, David H. Hughes\altaffilmark{8}, Itziar Aretxaga\altaffilmark{8}, Tracy Webb\altaffilmark{9}, Alejo Mart\'{i}nez--Sansigre\altaffilmark{10}, Sungeun Kim\altaffilmark{11}, Kimberly S. Scott\altaffilmark{3}, Jason Austermann\altaffilmark{3}, Thushara Perera\altaffilmark{3}, James D. Lowenthal\altaffilmark{12},  Eva Schinnerer\altaffilmark{10}, \& Vernesa Smol\v{c}i\'{c}\altaffilmark{10}}
\altaffiltext{1}{Harvard--Smithsonian Center for Astrophysics, 60 Garden Street, 
Cambridge, MA 02138}
\altaffiltext{2}{jyounger@cfa.harvard.edu}
\altaffiltext{3}{Astronomy Department, University of Massachusetts, Amherst, MA, 01003}
\altaffiltext{4}{Joint ALMA Office, El Golf 40, Las Condes, Santiago 7550108, Chile}
\altaffiltext{5}{National Astronomical Observatory of Japan (NAOJ), 2--21--1 Osawa, Mitaka, Tokyo 181--8588}
\altaffiltext{6}{Institute of Astronomy, University of Tokyo, 2--21--1, Osawa, Mitaka, Tokyo 181--0015}
\altaffiltext{7}{Nobeyama Radio Observatory, Minamimaki, Minamisaku, Nagano 384--1805}
\altaffiltext{8}{Instituto Nacional de Astrof\'{i}sica, \'{O}ptica y Electr\'{o}nica (INAOE), Tonantzintla, Peubla, M\'{e}xico}
\altaffiltext{9}{Department of Physics, McGill University, Rutherford Physics Building, 3600 rue University, Montr\'{e}al, Canada, H3A 2T8}
\altaffiltext{10}{Max--Planck--Institut f\"{u}r Astronomie, K\"{o}ningstuhl--17, D--69117, Heidelberg, Germany}
\altaffiltext{11}{Astronomy and Space Sciences Department, Sejong University, 98 Kwangjin--gu, Kunja--dong, Seoul, 143--747, Korea}
\altaffiltext{12} {Astronomy Department, Smith College,  Clark Science Center, Northampton, MA  01060} 
\altaffiltext{13} {Fellow of the International Max Planck Research School for Astronomy and Cosmic Physics} 

\begin{abstract}

We have used the Submillimeter Array to image a flux limited sample of seven submillimeter galaxies, selected by the AzTEC camera on the JCMT at 1.1 mm, in the COSMOS field at 890\micron with $\sim2\arcsec$ resolution.  All of the sources -- two radio--bright and five radio--dim -- are detected as single point--sources at high significance ($> 6\sigma$), with positions accurate to $\sim 0.2\arcsec$ that enable counterpart identification at other wavelengths observed with similarly high angular resolution.  All seven have IRAC counterparts, but only two have secure counterparts in deep HST/ACS imaging.  As compared to the two radio--bright sources in the sample, and those in previous studies, the five radio--dim sources in the sample (1) have systematically higher submillimeter--to--radio flux ratios, (2) have lower IRAC 3.6--8.0\micron fluxes, and (3) are not detected at 24\micronend.  These properties, combined with size constraints at 890\micron ($\theta \lsim 1.2\arcsec$), suggest that the radio--dim submillimeter galaxies represent a population of very dusty starbursts, with physical scales similar to local ultraluminous infrared galaxies, and an average redshift higher than radio--bright sources.
\end{abstract}


\section{Introduction}
Early studies of the far--infrared (FIR) cosmic background indicated that up to half of the cosmic energy density is generated by dusty starbursts and active galactic nuclei \citep{fixsen1998,pei1999}.  One of the most exciting developements of the past decade has been the resolution of a significant fraction of this background into discrete sources.  Deep, wide blank--field surveys at 850\micron \citep{smail1997,barger1998,hughes1998,eales1999,eales2000,cowie2002,scott2002,webb2003,serjeant2003,wang2004,coppin2006} with the Submillimeter Common--User Bolometric Array \citep[SCUBA;][]{holland1999} on the James Clerk Maxwell Telescope (JCMT), and later surveys at millimeter wavelengths  \citep{greve2004,dannerbauer2004,carilli2005,schlaerth2005,laurent2005,bertoldi2007}, revealed that this background was dominated by luminous (LIRG) and ultraluminous (ULIRG) infrared galaxies at high redshift $z\gsim 2$.  Multi--wavelength follow--up studies of these submillimeter galaxies (SMGs) showed that they are massive, young objects seen during their formation epoch, with very high specific star formation rates that may account for up to $\sim50$\% of the cosmic star formation at $z>1$ \citep[see review by][]{blain2002}.

Progress towards a thorough understanding of the physical processes driving SMGs has been hampered by two factors: their faintness at optical wavelengths, and the relatively poor ($\gsim 10\arcsec$) angular resolution of the current generation of submillimeter cameras.  The first significant breakthrough came with deep radio surveys, which found a correlation between submillimeter and radio continuum emission \citep{ivison1998,chapman2001,ivison2002,dunlop2004,ivison2007}.  This localized SMGs to a few tenth's of an arcsecond, and allowed the first spectroscopic observations \citep{chapman2003a,chapman2005} which showed that SMGs lie at high redshift $2\lsim z \lsim 3$ with a median of $z \sim 2.5$.  However, only a fraction of these redshifts have been confirmed via CO \citep{greve2005,tacconi2006} or mid--infrared \citep{lutz2005,menendez2007,valiante2007} spectroscopy.  Furthermore, the rapid dimming of the radio continuum with redshift \citep[$I\sim (1+z)^{-(4+\alpha)}$, $\alpha=0.8$;][]{condon1992} means existing radio--confirmed SMG samples, which represent $\sim3/4$ of the overall SMG population \citep[e.g.,][]{ivison2002}, are relatively insensitive to systems at $z\gsim 3$, and thus are biased.  Recent studies have suggested that near to mid--infrared imaging using the Infrared Array Camera \citep[IRAC:][]{fazio2004}, in combination with 24\micron observations using the Multiband Imaging Photometer \citep[MIPS:][]{rieke2004}, on board the {\it Spitzer Space Telescope} may offer an alternative to radio identification.  However, this technique relies on either purely statistical arguments \citep{pope2006} or broad band infrared color criteria \citep{ashby2006}, and also may be subject to biases.

Reliable counterpart identification represents potentially the most challenging obstacle to a more complete understanding of SMGs.  Previous interferometric observations at millimeter \citep{downes1999,frayer2000,dannerbauer2002,downes2003,genzel2003,kneib2005,greve2005,tacconi2006} and submillimeter \citep{iono2006} wavelengths have identified unambiguous counterparts for increasing numbers of radio--detected SMGs, and have confirmed the radio--submillimeter association.  However, to date there has been no reliable high--resolution followup of a uniformly selected sample including radio--undetected SMGs, and the true nature of these sources therefore remains elusive.  

In this work, we present high--resolution 890\micron interferometric imaging by the Submillimeter Array \citep[SMA:][]{ho2004} of a flux--limited sample of  sources selected at 1.1~mm by the AzTEC Camera \citep{wilson2007} on the JCMT, in a survey of a section of the COSMOS field \citep{scott2007}.  The SMA has confirmed all seven of the AzTEC targets at arcsecond resolution, with positions accurate to $\sim 0.2\arcsec$.  In \S~\ref{sec:obs} we describe our observations, and in \S~\ref{sec:astrometry} we address sources of uncertainty in the derived positions.  In \S~\ref{sec:individual} we describe each of the sources, and in \S~\ref{sec:discussion} we discuss some potential interpretations of the data.  All magnitudes are given in the AB system \citep{oke1974}.

\section{Observations and Data Reduction}
\label{sec:obs}

\begin{deluxetable*}{cccccc}
\tablecaption{Astrometry of SMA/AzTEC Sources}
\tablehead{
\colhead{} & \colhead{Name} & \colhead{$\sigma(\alpha)$} & \colhead{$\sigma(\delta)$} & 
\colhead{AzTEC Offset} & \colhead{IRAC Offset$^a$}
}
\startdata
AzTEC1 & AzTEC J095942.86+022938.2 & 0.11\arcsec & 0.20\arcsec & 3.3\arcsec & 0.3\arcsec \\
AzTEC2 & AzTEC J100008.05+022612.2 & 0.13\arcsec & 0.23\arcsec & 0.3\arcsec & \nodata$^b$ \\
AzTEC3 & AzTEC J100020.70+023520.5 & 0.19\arcsec & 0.31\arcsec & 1.6\arcsec & 0.9\arcsec \\
AzTEC4 & AzTEC J095931.72+023044.0 & 0.15\arcsec & 0.24\arcsec & 3.5\arcsec & 0.5\arcsec \\
AzTEC5 & AzTEC J100019.75+023204.4 & 0.16\arcsec & 0.11\arcsec & 1.7\arcsec & 0.8\arcsec \\
AzTEC6 & AzTEC J100006.50+023837.7 & 0.19\arcsec & 0.28\arcsec & 2.8\arcsec & 0.7\arcsec \\
AzTEC7 & AzTEC J100018.06+024830.5 & 0.24\arcsec & 0.29\arcsec & 1.5\arcsec & 0.5\arcsec \\
\enddata
\label{tab:offset}
\tablenotetext{a}{Relative to IRAC Channel 1 source, which has an astrometric uncertainty of $\sim0.2\arcsec$ and an angular resolution of $\sim1.6\arcsec$.\\}
\tablenotetext{b}{IRAC counterpart is confused with a bright foreground object.\\}
\end{deluxetable*}

The COSMOS field \citep{scoville2006} benefits from an extraordinary wealth of deep, multi--wavelength coverage from the X--ray to the radio.  In this work, we utilize $i$ band imaging with the Advanced Camera for Surveys \citep[ACS:][]{ford1998} on board the {\it Hubble Space Telescope} to a depth of 27.1 magnitudes \citep{koekemoer2007}, a variety of ground--based optical and near--infrared imaging data \citep[see ][]{taniguchi2006,capak2007}, IRAC and MIPS imaging at 3.6, 4.5, 5.8, 8.0, and 24\micron to $5\sigma$ depths of $\sim$0.9, 1.7, 11.3, 14.6, and 71\microjy respectively \citep{sanders2007}, and 1.4 GHz radio continuum imaging to a mean rms depth of $\sim 10.5$\microjyend/beam with the Very Large Array \citep[VLA:][]{schinnerrer2006}.

The AzTEC/COSMOS survey covers 0.15 deg$^2$ of the COSMOS field at 1.1 mm with an rms noise level of 1.3 mJy/beam \citep{scott2007}.  The AzTEC/COSMOS catalog includes 44 sources with $S/N \ge 3.5\sigma$ and 10 robust sources with $S/N\ge5\sigma$.  For our SMA observations we chose the seven highest significance sources, which effectively yielded a flux--limited sample of millimeter selected SMGs.

Five of these sources (AzTEC1--4 and AzTEC6) have either weak ($F_{20cm} < 60$\microjyend) or no radio sources within the AzTEC beam ($18\arcsec$ FWHM), which we designate {\em radio--dim}.  The remaining two sources (AzTEC5 and AzTEC7), with strong radio sources ($F_{20cm} = 161$ \& 196 \microjyend) within the AzTEC beam are designated {\em radio--bright.}   This convention was chosen to address the inherent ambiguity of the radio detected versus undetected designation often used in the literature; as \citet{pope2006} observed, SMGs without a radio counterpart likely do not represent a distinct population, but rather lie just below the detection threshold for a given survey.

The SMA observations were performed in the compact array configuration (beam size $\sim 2\arcsec$) at 345 GHz (full bandwidth 2 GHz) from January through March 2007.  The weather was excellent, with typical rms noise levels of 1.0--1.5 mJy per track with $\sim 6$ hours of on--source integration.   The data were calibrated using the {\sc mir} software package \citep{scoville1993}, modified for the SMA.  Complex gain calibration was performed using the calibrator sources  J1058+015 ($\sim 3$ Jy, $\sim15^\circ$ away from targets) and J0854+201 ($\sim 1$ Jy, $\sim24^\circ$ away from targets).  Passband calibration was done using available strong calibrator sources, primarily 3C273 and Callisto. The absolute flux scale was set using observations of Callisto and is estimated to be accurate to better than 20\%.  Positions and fluxes of the COSMOS sources were derived from the calibrated visibilities using the {\sc miriad} software package \citep{sault1995}.

\section{Astrometric Uncertainties from Interferometric Imaging}
\label{sec:astrometry}

Precise astrometry is one of the most valuable contributions of interferometric observations to the study of SMGs, and accurate characterization of the positional uncertainty is crucial. There are two factors to consider when estimating astrometric accuracy of SMA observations of SMGs: (1) statistical errors due to noise in fitting a point--source to the calibrated visibilities, and (2) systematic errors due to uncertainties in the interferometer baselines \citep[see e.g.,][]{downes1999}. In general, the 1--D statistical uncertainty in position scales as $\sim 0.5 \theta/(S/N)$, where $(S/N)$ is the signal--to--noise of the fit and $\theta$ is the FWHM of the beam \citep{reid1988}. This expectation is borne out in the {\sc miriad} fitting routines, which for the $\sim2\arcsec$ FWHM SMA beam and $(S/N)\approx10$ yield typical uncertainties of $\sim 0.1 \arcsec$ in $\alpha$ and $\delta$. Systematic uncertainties, or errors related to uncertainties in the baselines, scale as $\sim A (\Delta s/\lambda) R \theta$, where $\Delta s$ is the baseline error, $R$ is the distance from the calibrator with known position in radians, and $A$ is a constant of order unity that is sensitive to the details of the array--source geometry. For the SMA compact array configuration, the baseline parameters are typically measured to better than 0.1~millimeters rms.  To obtain an empirical upper limit on the systematic position error induced by baseline errors, we use one of the calibrators, J1058+015, to calibrate the other, J0854+201 (35 degrees away, more than twice the distance of J1058+015 to the COSMOS field), and examine the resulting offsets.\footnote{The calibrator sources are sufficiently strong that statistical errors are insignificant.} This procedure yields a typical systematic total angular offset of J0854+201 of $\sim 0.2\arcsec$ from its known position. We combine this conservative estimate of the systematic uncertainties with the measured statistical error, in the uncertainties listed in Table~\ref{tab:offset}.

\section{Notes on Individual Objects}
\begin{deluxetable*}{cccccccccccccc}
\tablewidth{0pt}
\tablecaption{Photometry of SMA/AzTEC Sources}
\tablehead{
\colhead{} & \colhead{$F_{1100\mu m}$} & \colhead{$F_{890\mu m}$} & 
\colhead{$F_{3.6\mu m}^a$} & \colhead{$F_{4.5\mu m}^a$} & \colhead{$F_{5.8\mu m}^a$} & 
\colhead{$F_{8.0\mu m}^a$} & \colhead{$F_{24\mu m}^b$} & \colhead{$F_{20cm}^c$} \\
\colhead{} & \colhead{(mJy)} & \colhead{(mJy)}  & \colhead{($\mu$Jy)} &
\colhead{($\mu$Jy)} & \colhead{($\mu$Jy)} & \colhead{($\mu$Jy)} & 
\colhead{($\mu$Jy)} & \colhead{($\mu$Jy)}
}
\startdata
AzTEC1& $10.7\pm1.3$ & $15.6\pm 1.1$ & $4.6\pm1.0$ & $4.6\pm1.4$ & $<11.2$ & $17.6\pm8.1$ & $<71$ & $48\pm14$ \\
AzTEC2$^d$ & $9.0\pm1.3$ & $12.4\pm1.0$ & \nodata & \nodata & \nodata & \nodata & \nodata & $52\pm14$ \\
AzTEC3 & $7.6\pm1.2$ & $8.7\pm1.5$ & $3.9\pm1.0$ & $3.6\pm1.4$ & $<11.2$ & $<13.4$ & $<71$ & $<41$ \\
AzTEC4 & $6.8\pm 1.3$ & $14.4\pm1.9$ & $4.8\pm1.0$ & $5.1\pm1.4$ & $12.4\pm6.7$ & $19.6\pm8.1$ & $<71$ & $<41$ \\
AzTEC5 & $7.6\pm1.3$ & $9.3\pm1.3$ & $8.8\pm1.0$ & $9.0\pm1.4$ & $11.6\pm6.7$ & $32.7\pm8.1$ & $164\pm20$ & $161\pm35$ \\
AzTEC6 & $7.9\pm1.2$ & $8.6\pm1.3$ & $2.4\pm1.0$ & $2.4\pm1.4$ & $<11.2$ & $<13.4$ & $<71$ & $<41$ \\
AzTEC7 & $8.3\pm1.4$ & $12.0\pm1.5$ & $52.1\pm1.0$ & $52.1\pm1.4$ & $80.6\pm6.7$ & $63.4\pm8.1$ & $550\pm20$ & $196\pm61$ \\
\enddata
\tablenotetext{a}{Fluxes are measured in a $3\arcsec$ aperture.  Errors and flux limits are the $3\sigma$ rms and $5\sigma$ rms fluctuation within that aperture respectively.  Aperture corrections were done to the IRAC calibration radius of 12$\arcsec$. \\}
\tablenotetext{b}{Fluxes were measured in a $5\arcsec$ aperture.  Errors are the $3\sigma$ rms fluctuations in that aperture, and flux limits are the $5\sigma$ sensitivity from \citet{sanders2007}.  Aperture corrections were done to the MIPS calibration radius of 35$\arcsec$.\\}
\tablenotetext{c}{$F_{20cm}$ measurements have been corrected by a factor of 1.15-1.2 for bandwidth smearing \citep[see][]{bondi2007}. Flux limits are at $3\sigma$.\\}
\tablenotetext{d}{IRAC and MIPS are confused with a bright foreground object.\\}
\label{tab:params}
\end{deluxetable*}

\begin{figure*}
\plotone{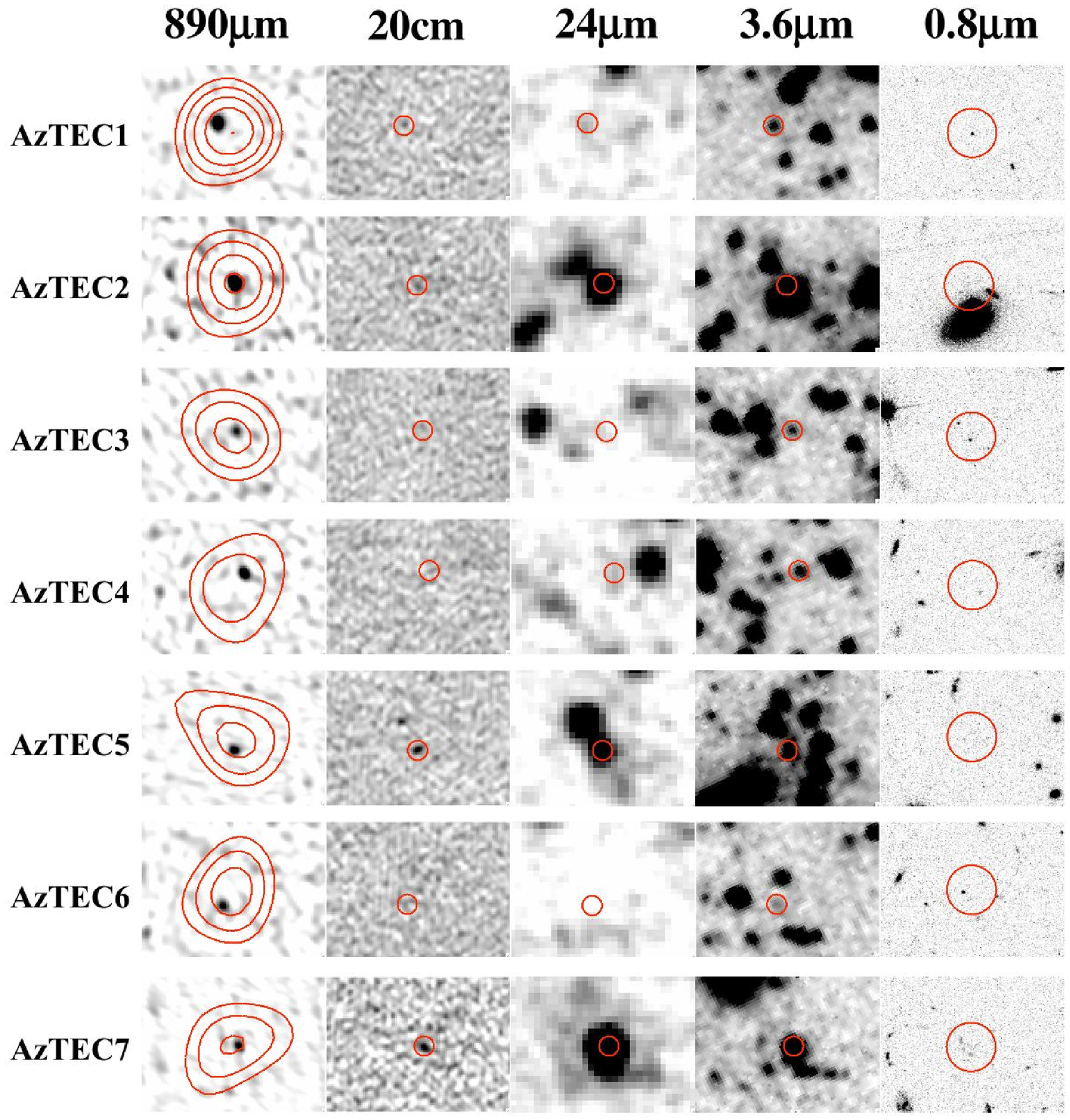}
\caption{Stamp images for the SMA/AzTEC sources (top to bottom AzTEC1--7) for (left to right) SMA (890\micronend), VLA radio continuum (20cm), MIPS Channel 1 (24\micronend), IRAC Channel 1 (3.6\micronend), and ACS ($i$--band; 0.8\micronend) imaging data.  Overlayed in red on the SMA image are contours at 3$\sigma$, 4$\sigma$,... from AzTEC imaging data \citep{scott2007}.  The red circles in the remaining stamps have a radius $2\arcsec$, corresponding to twice the FWHM of the SMA beam, at the SMA position.  Two sources (AzTEC 1 \& 7) have secure optical counterparts in the ACS images, while AzTEC 3 \& 6 have candidate optical counterparts that are a potential foreground object (see \S~\ref{sec:individual}) and outside the SMA beam respectively.  Each stamp image is $37\arcsec\times 27\arcsec$, with the exception of the ACS stamps which are $15\arcsec\times 11\arcsec$.}
\label{fig:stamps}
\end{figure*}

\label{sec:individual}
Astrometry and photometry for all the targets are included in Table~\ref{tab:offset} and \ref{tab:params} respectively, and postage stamps are shown in Figure~\ref{fig:stamps}.  Here we comment on the individual objects. \\

{\it AzTEC J095942.9+022938.2} (AzTEC1) -- AzTEC1 is the brightest submillimeter source in our sample, and is detected at 14$\sigma$ significance by the SMA.   A weak radio source ($F_{20cm} = 48\pm14$\microjyend) is coincident with the SMA position.  There are also IRAC 3.6, 4.5, and 8.0\micron sources coincident with the SMA position, but no significant MIPS 24\micron emission.  There is a compact B--band dropout offset from the SMA position by $0.1\arcsec$ with $i = 25.25\pm 0.79$ mag in the ACS mosaic which we believe is the optical counterpart.   The B--band dropout nature of this source  suggests that it lies at $3.5\lsim z \lsim 4.5$ \citep[see e.g.,][]{steidel1999,giavalisco2004}, which is consistent with the lack of strong radio or 24\micron emission expected from a starburst galaxy at that redshift.    \\

{\it AzTEC J100008.0+022612.2} (AzTEC2) -- AzTEC2 is detected at 12$\sigma$ significance by the SMA.  There is a weak radio source ($F_{20cm} = 52\pm14$\microjyend) coincident with the SMA position, but it has no candidate optical counterpart.  ACS imaging reveals that it is offset by $\sim 3\arcsec$ from a foreground galaxy that is very bright in IRAC and MIPS.  Even a careful subtraction, using the ACS data convolved with IRAC and MIPS point spread functions to remove the foreground object, does not reveal a potential counterpart; AzTEC2 is either not detected in the near and mid infrared, or is severely confused with a foreground galaxy that is not associated with the submillimeter continuum emission. \\

{\it AzTEC J100020.7+023520.5} (AzTEC3) -- AzTEC3 is detected at 6$\sigma$ by the SMA.  There are IRAC detections at 3.6 and 4.5\micron coincident with the SMA position, but no likely MIPS 24\micron or radio counterparts.  There is an optical detection with $i=25.91\pm 1.07$ in the ACS mosaic within $0.3\arcsec$ of the SMA position.  The lack of 24\micron emission suggests that the source is at $z\gsim 3$, or is at $z\sim 1.5$ and has a deep rest--frame 9.7\micron silicate absorption feature.  From the observed 1.1 mm flux, and assuming a grey--body \citep[see][]{yun2002} with $T_d\approx 40-50$K and $\beta\approx 1.5-2$ and the local FIR--radio correlation of \citet{condon1992}, we would expect strong 20cm counterparts with $F_{20cm}\gsim 100$\microjy at $z\sim 1.5$.  Therefore if this optical source is at lower redshift, it either has lower radio emission than would be expected from the local FIR--radio relation, or is a chance alignment and not the correct counterpart. \\

{\it AzTEC J095931.7+023044.0} (AzTEC4) -- AzTEC4 is detected at 7$\sigma$ by the SMA. It has no candidate optical, radio, or 24\micron counterparts, but is detected by IRAC at 3.6, 4.5, 5.8, and 8.0\micronend.  \\

{\it AzTEC J100019.8+023204.2} (AzTEC5) -- AzTEC5 is detected at 8$\sigma$ by the SMA.  It has a radio counterpart coincident with the SMA position, and is detected by IRAC at 3.6, 4.5, 5.8, and 8.0\micronend, and by MIPS at 24\micronend.  There are two significant radio sources within the AzTEC beam, with fluxes of $F_{20cm} = 161\pm35$ and $81\pm12$ \microjyend.  One of these radio sources is singled out as the counterpart by the high angular resolution SMA imaging.  Furthermore, the correct 24\micron counterpart is the weaker of the two within the AzTEC beam.  There are, however, no associated optical sources in the ACS $i$ band image.   The IRAC fluxes follow a power--law, which is consistent with a very dusty active galactic nucleus (AGN).  \\

{\it AzTEC J100006.5+023837.7} (AzTEC6) -- AzTEC6 is detected at 6.5$\sigma$ by the SMA. It has no candidate optical, radio, or 24\micron counterparts, but is detected by IRAC at 3.6 and 4.5\micronend.  \\

{\it AzTEC 100018.1+024830.5} (AzTEC7)--  AzTEC7 is detected at 8$\sigma$ by the SMA.   Like AzTEC5, there is a radio counterpart ($F_{20cm} = 196\pm61$\microjyend), and it is detected by IRAC at 3.6, 4.5, 5.8, and 8.0\micron, and by MIPS at 24\micronend.  Its SED peaks at 5.8\micronend, and is very bright at 24\micronend, which is consistent with a $z\sim 2.5$ starburst.  There is also an optical counterpart in the ACS imaging with a disturbed morphology, reminiscent of a merging system.  \\
\newpage
\section{Discussion}

\label{sec:discussion}

Astrometry with the SMA highlights the unique power of this instrument; secure multi--wavelength counterparts for many of the targets could only be identified via interferometric imaging.  We find that, while there is always -- with the exception of the highly confused case of AzTEC2 -- an IRAC counterpart coincident with the SMA position, there are often several 24\micron sources within the AzTEC beam (see Figure~\ref{fig:stamps}) that are not associated with the submillimeter emission.  This is contrary to the prevailing wisdom, in which radio--dim SMGs, like their radio--bright counterparts, are associated with redshifted strong polycyclic aromatic hydrocarbon (PAH) emission features in the 24\micron band.  For {\em all five} of the radio--dim sources (AzTEC1--4 and AzTEC6) there are proximate 24\micron sources that, if selected, would lead to misidentification of multiwavelength counterparts to the submillimeter source (see Figure~\ref{fig:stamps}).  AzTEC5 has two potential radio counterparts within the AzTEC beam, both of which have strong 24\micron emission.  In this case as well, only the SMA can unambiguously identify the correct counterpart.

\begin{figure}
\plotone{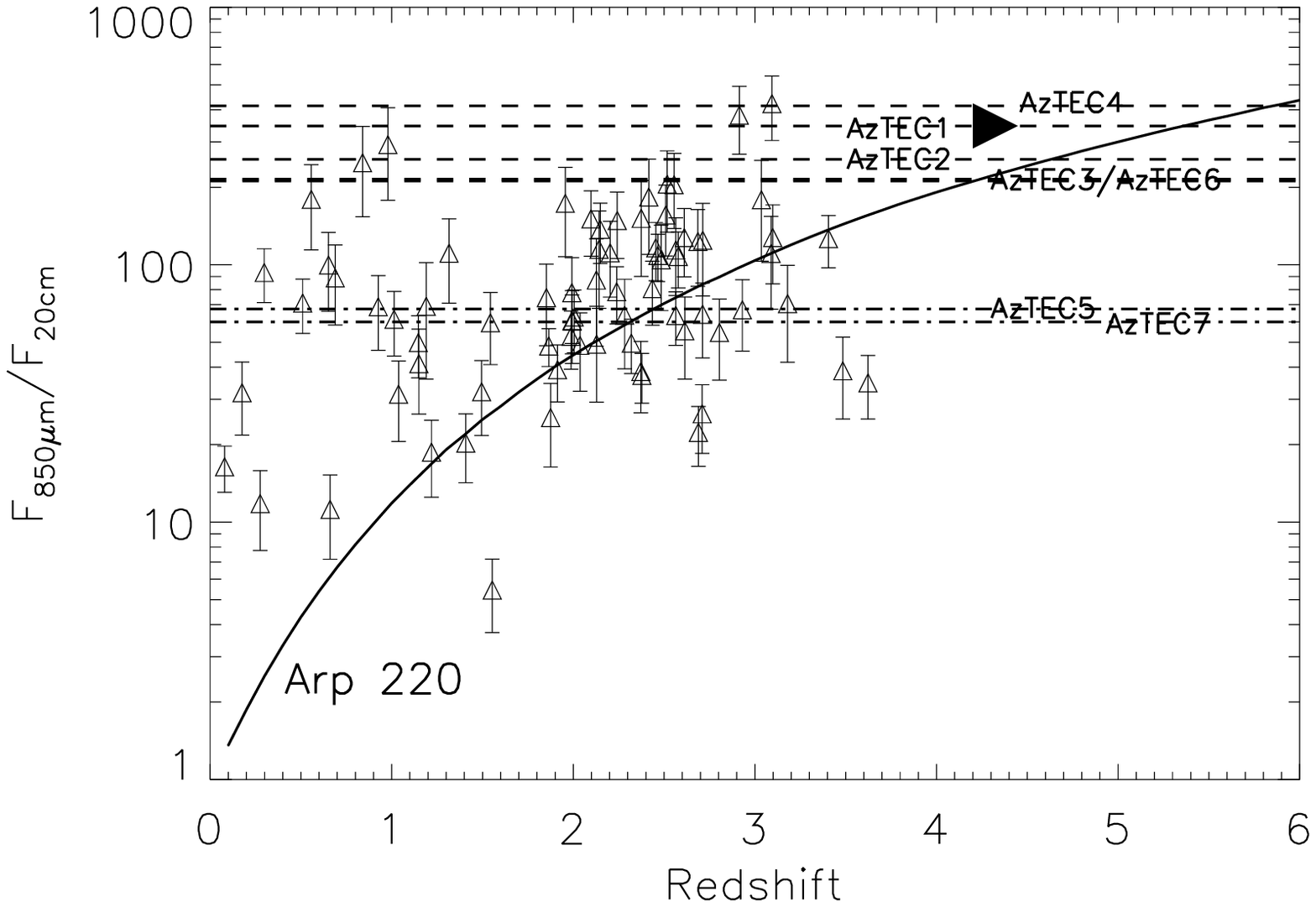}
\caption{The change with redshift of the ratio of the submillimeter (850\micronend) versus radio (20cm) continuum emission in SMGs \citep[see][]{carilli1999}.  Radio--bright SMGs with optical spectroscopic redshifts from \citet{chapman2005} are shown as open triangles, as compared to radio--dim (dashed lines) and radio--bright (dash--dot lines) SMA/AzTEC sources (see \S~\ref{sec:individual} for abbreviations).  The filled triangle represents a rough redshift of $z\gsim 4$ for AzTEC1 from the B--band dropout nature of its optical counterpart.  The solid line is a model track for Arp 220.  SMA flux measurements at 890\micron were corrected to 850\micron using the $F_{890\mu m}/F_{1100\mu m}$ ratio from SMA and AzTEC, a $\lsim15\%$ correction for all the sources.}
\label{fig:chapman_radio}
\end{figure}

Only now, with these counterparts, can we look for clues to the nature of the radio--dim SMG population.  It has been suggested that the ratio of the submillimeter to radio flux ($F_{850\mu m}/F_{20cm}$) is a potentially useful redshift indicator \citep{carilli1999,yun2002,aretxaga2007}.  In Figure~\ref{fig:chapman_radio}, we plot this ratio for both the SMA/AzTEC sources and for radio--bright SMGs with optical spectroscopic redshifts \citep[C05:][]{chapman2005}.   The location of the radio--dim sources above the locus of points from C05 is, as previous authors have speculated \citep[e.g.,][]{carilli1999,chapman2005,pope2006,ivison2007}, consistent with either colder dust temperatures or a higher average/median redshift than the C05 sample.

\begin{figure}
\plotone{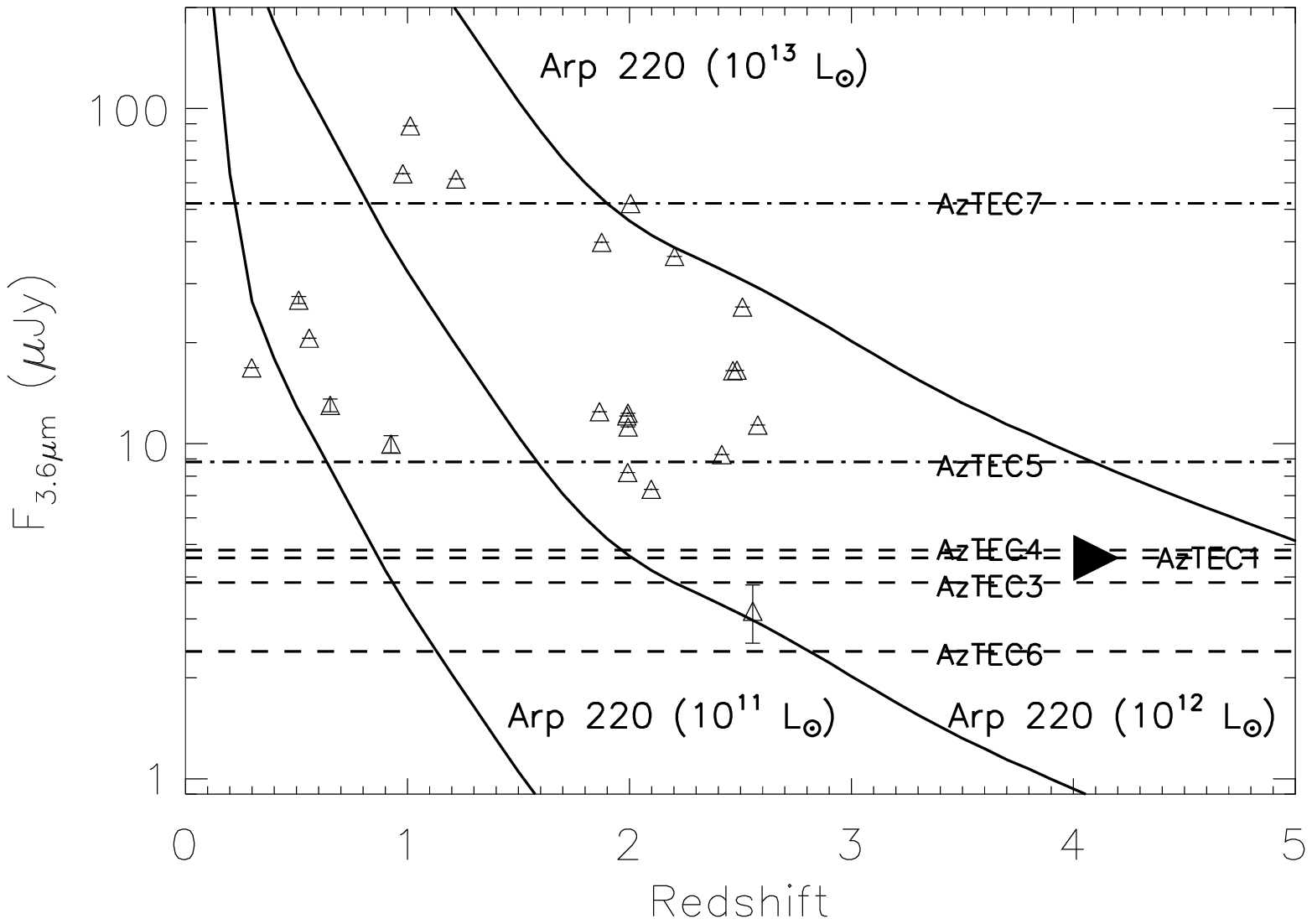}
\caption{IRAC 3.6\micron fluxes versus redshift for radio--bright SMGs with optical redshifts \citep[open triangles;][]{chapman2005}, as compared to radio--dim (dashed lines) and radio--bright (dash--dot lines) SMA/AzTEC sources (see \S~\ref{sec:individual} for abbreviations).  The filled triangle represents a rough redshift of $z\gsim 4$ for AzTEC1 from the B--band dropout nature of its optical counterpart.  For comparison, we include model tracks for Arp 220 with total luminosities of $10^{11} L{\odot}$ (a LIRG), $10^{12} L{\odot}$ (a ULIRG), and $10^{13} L_{\odot}$ (a HyLIRG).}
\label{fig:chapman}
\end{figure} 

 The IRAC and MIPS counterparts for the radio--dim SMA/AzTEC sources support this hypothesis.  Given the 1.1 mm selection function and cosmic volume probed, we would expect a sample with $F_{1100\mu m}>6.5$ mJy to be dominated by objects at $z\gsim 1$ with total infrared luminosities on the order of $10^{12-13} L_{\odot}$.  For such systems, the observed 3.6\micron flux should be a strong function of redshift.  In Figure~\ref{fig:chapman}, we show 3.6\micron fluxes for the SMA/AzTEC sources, compared to C05 sources in the Hubble Deep Field North (HDFN) and SSA22.  The radio--dim sources have systematically lower fluxes which, assuming an Arp 220 model, suggests that they may lie at higher redshift.  Furthermore, these same radio--dim sources do not have 24\micron detections, indicating that either the 7.7\micron PAH emission features have been redshifted out of, or the 9.7\micron silicate absorption feature is in the 24\micron MIPS band, which suggest either $z\gsim 3$ or $z\sim 1.5$ respectively.  However, as a caveat we note that it is not impossible that they are intrinsically different objects than Arp 220, with fainter PAHs or different submillimeter--to--radio ratios.  At the same time, radio--bright SMA/AzTEC sources are consistent with the population observed by C05.  

Therefore, the infrared and radio properties of radio--dim SMA/AzTEC sources in a 1.1~mm flux--limited sample suggest that they lie at high redshift $z\gsim 3$.  Furthermore, their ubiquity in both this (70\% of sources with $S/N > 5$ at 1.1~mm) and other \citep[50\% of sources with $S/N > 4$ at 1.2~mm;][]{bertoldi2007} millimeter surveys further suggest that they contribute significantly to the observed millimeter/submillimeter number counts, and that the median redshift of $z\sim2.5$ inferred by C05 using assumed radio counterparts is a lower limit.  A detailed discussion of the properties of these systems awaits a full SED analysis, which we postpone to a future paper.  However, assuming an Arp 220 model, the observed submillimeter flux of these sources implies very high total infrared luminosities of $L(8-1000\mu m) \gsim 10^{13}\, L_{\odot}$ -- or $\gsim 5\times 10^{12}\, L_{\odot}$ assuming a Mrk 231 model -- which is similar to high--redshift hyperluminous infrared galaxies \citep[HyLIRGs:][]{huang2006}.  The presence of a significant population of these objects has important consequences for models of hierarchical galaxy formation, which are only beginning to account for such systems at later epochs \citep[$z\sim2$;][]{baugh2005}.   It is also curious that a majority (60 or 70\%) of the most luminous sources in the AzTEC/COSMOS catalog are radio--dim; as first noted by \citet{ivison2002}, the most luminous SMGs may have a higher average/median redshift.

Furthermore, such a population of massive, dusty starbursts at $z\gsim 3$ constrains models of dust production, given the limited look--back time since the formation of the first stars at $z\approx 20-30$ \citep[][and references therein]{bromm2004}.  The dust mass corresponding to the observed thermal emission is approximately equal to $M_d \approx L_\nu/4\pi\kappa_\nu B_\nu(T_d)$, where $L_\nu$ is the observed luminosity at a given rest--frame frequency $\nu$, $\kappa_\nu$ is the dust opacity at that frequency, and $B_\nu(T_d)$ is the blackbody emission at the effective dust temperature $T_d$.  For redshifts of $z\gsim 3$, assuming the \citet{weingartner2001} Milky Way dust opacity, dust temperature $T_d = 45-70$ K, and a flat $\Lambda$CDM cosmology, we find that the observed 345 GHz (rest--frame $> 1380$ GHz) flux of our objects ($F_{890\mu m} \approx 10$ mJy) implies dust masses of order $0.6-3\times 10^9\, M_{\odot}$ at the observed time.   If dust production is dominated by evolved, post--main sequence stars with ages $\gsim 1$ Gyr, as it is locally \citep{gehrz1989,marchenko2006}, this requires a dust production rate of $\dot{M_d} \gsim 0.7-3.4\, M_{\odot}$ yr$^{-1}$ over the same dust temperature range.  Or, if supernovae are also significant contributors to dust production at high redshift \citep[e.g.,][]{dunne2003}, as may be the case in high redshift quasars \citep{maiolino2004}, then the required rate of production is lower by $\sim 50\%$.

Finally, SMA imaging in combination with a redshift constraint allows us to place limits on the spatial extent of the submillimeter continuum.  All seven of the sources are compact single sources; the real visibility amplitudes indicate that they are unresolved out to the longest baselines, from which we infer a maximum angular size of $\sim 1.2\arcsec$.  This is particularly interesting for this sample of the brightest sources in the AzTEC/COSMOS catalog because it rules out blends of multiple fainter sources as a significant contributor to the upper end of the observed SMG luminosity function.  It also agrees with previous interferometric measurements of the angular extent of the millimeter \citep{downes1999,genzel2003,downes2003,kneib2005,tacconi2006} and submillimeter \citep{iono2006} emission from SMGs, and marginally with those of the radio continuum \citep{chapman2004}.  Assuming a flat $\Lambda$CDM cosmology, these angular constraints correspond to a physical scale for the submillimeter continuum of 10 \lengthunits at $z\sim 2$ and 8 \lengthunits at $z\sim 4$.  These size scales are consistent with far--infrared continuum emission associated with a merger driven starburst \citep[e.g.,][]{mihos1994} analogous to local luminous and ultraluminous infrared galaxies \citep{downes1998,sakamato1999,sakamato2006,iono2007}, and are potentially in conflict with cool extended cirrus dust models \citep{efstathiou2003,kaviani2003} and a monolithic collapse scenario.

\section{Conclusion}
\label{sec:conclusion}

We use the SMA to follow--up the brightest millimeter sources in the AzTEC/COSMOS survey \citep{scott2007}.  All seven sources, including five radio--dim ($F_{20cm} < 60$\microjyend) SMGs, are detected at high significance ($> 6\sigma$) with derived positions accurate to $\sim 0.2\arcsec$.  All seven of the sources, with the possible exception of one highly confused case, are coincident with IRAC detections, and all but two are optical dropouts.  We find that the radio--dim SMGs in our sample have systematically higher submillimeter--to--radio ratios and lower IRAC 3.6--8.0\micron fluxes than radio--bright sources, and are not detected at 24\micronend.  This, in combination with size constrains from the imaging data, suggests that radio--dim SMGs represent a population of very dusty $z\gsim 3$ starbursts with physical scales similar to local ULIRGs.

\acknowledgements

Thanks to the anonymous referee for comments.  We also thank Olivier Ilbert for his help in estimating photometric redshifts, and Herve Aussel, Dave Sanders, and the rest of the COSMOS-Spitzer team for providing the reduced IRAC and MIPS data.  We also thank Lars Hernquist, Phillip F. Hopkins, T. J. Cox, R. J. Ivison, Ian Smail, Yuexing Li, Du\v{s}an Kere\v{s}, and Sukanya Chakrabarti for helpful discussion. The Submillimeter Array is a joint project between the Smithsonian Astrophysical Observatory and the Academia Sinica Institute of Astronomy and Astrophysics and is funded by the Smithsonian Institution and the Academia Sinica.  This work is based on observations made with the {\it Spitzer} Space Telescope, which is operated by the Jet Propulsion Laboratory, California Institute of Technology, under NASA contact 1407.  This work is partially funded by NSF Grant \#0540852.  The JCMT/AzTEC Survey and S. Kim were supported in part by the Korea Science and Engineering Foundation (KOSEF) under a cooperative agreement with the Astrophysical Research Center of the Structure and Evolution of the Cosmos (ARCSEC).  DHH and IA are supported in part by CONACyT.

{\em Facilities:} \facility{SMA}, \facility{JCMT}, \facility{Spitzer (IRAC, MIPS)}, \facility{HST (ACS)}, \facility{VLA}

\bibliographystyle{apj}
\bibliography{../../smg}

\clearpage

\end{document}